\documentclass[10pt,preprint2]{aastex}

\usepackage{bm}

\shorttitle{Pioneer anomally}
\shortauthors{Hodge}

\begin{document}

\title{Comments on ``The Pioneer Anomaly in the Light of New Data''}
\author{John C. Hodge}
\affil{Blue Ridge Community College, 100 College Dr., Flat Rock, NC, 28731-1690, jch9496@blueridge.edu \\*
\vspace{5mm}
}

\maketitle

\citet{tury09} reported on the status of the analysis of recently recovered Pioneer 10 (P10) and Pioneer 11 (P11) flight data and commented on some, but not all, of the characteristics of the Pioneer Anomaly (PA) that {\it{must}} be explained by a candidate model. {\it{Only one}} model presented to date is consistent with {\it{all}} the characteristics. 
\vspace{5mm}

The characteristics discussed by \citet{tury09} are: (i) that the anomalous data are interpreted, but unproven, to be sunward directed and to be an acceleration $a_\mathrm{P}$, (ii) that the more distant values of $a_\mathrm{P} \approx cH_o$ and are a blueshift rather than a redshift, (iii) that the Saturn and Jupiter encounter data are significantly different than later data, (iv) that the data analysis before the encounters was insufficient to detect the PA rather than there was no anomaly at less than 10 AU, (v) that the direction of the PA may be other than sun directed, and (vi) that analysis of individual spacecraft data indicates a possible difference in the $a_\mathrm{P}$ values. Additional PA characteristics that must also be explained by a model are: (vii) that the PA has an annual periodicity \citep{ande02}, (viii) that the PA has an Earth sidereal daily periodicity, (ix) that the $a_\mathrm{P}$ calculation by the {\textit{Sigma}} and CHASMP program methods for P10 (I) and P10 (II) show a discrepancy while showing consistency for P10 (III) \citep[Table I]{ande02}, (x) that the $a_\mathrm{P}$ of both spacecraft may be declining with distance \citep[as shown by the envelope in Fig.~1]{tury}, (xi) that the value of $a_\mathrm{P}$ averaged over a period during and after the Saturn encounter had a relatively high uncertainty \citep{niet05} that may be interpreted as high variability over the measurement duration, and (xii) that a modification of gravity large enough to explain the PA is inconsistent with the planetary ephemeredes unless the Equivalence Principle is violated \citep{ande02}. 

Further, (iii), (vi), (vii), (viii), (ix), (x), and (xi) suggest the PA is variable and environment dependent rather than is a fixed value. Also, (vii) and (viii) suggest, but not prove, the PA is Earth directed.

That the PA is an acceleration is unproven. The PA is measured by an unexplained frequency blueshift in the radio signal. The `` acceleration'' nomenclature is based on the unsupported hypothesis that the frequency shift is a Doppler effect. Other phenomena cause frequency shifts such as gravity using the Weak Equivalence Principle as shown in the Pound-Rebka experiment \citep{pound}. 

\citet{bert} concluded a scalar field is able to explain the PA. A scalar potential model \citep{hodg} is consistent with {\it{all}} the PA characteristics including a cosmological connection and variable $a_\mathrm{P}$.

\end{document}